\documentclass{singlecol}
\pdfoutput=1

\usepackage[T1]{fontenc}
\usepackage[pdftex]{graphicx}
\usepackage[caption=false,font=footnotesize]{subfig}

\usepackage{natbib}

\begin{document}

\setcounter{page}{1}

\LRH{J. Hauffa et~al.}

\RRH{An evaluation of keyword extraction from online communication}

\subtitle{}

\title{An evaluation of keyword extraction from online communication for the characterisation of social relations}

\authorA{Jan Hauffa, Tobias Lichtenberg, Georg Groh}
\affA{Department of Informatics,\\ Technische Universit\"at M\"unchen,\\ Boltzmannstr. 3, 85748 Garching, Germany \\
E-mail: \{hauffa,lichtent,grohg\}@in.tum.de}

\begin{abstract}
The set of interpersonal relationships on a social network service or a similar online community is usually highly heterogenous. The concept of tie strength captures only one aspect of this heterogeneity. Since the unstructured text content of online communication artefacts is a salient source of information about a social relationship, we investigate the utility of keywords extracted from the message body as a representation of the relationship's characteristics as reflected by the conversation topics. Keyword extraction is performed using standard natural language processing methods. Communication data and human assessments of the extracted keywords are obtained from Facebook users via a custom application. The overall positive quality assessment provides evidence that the keywords indeed convey relevant information about the relationship. \\
{\em This article is a revised and expanded version of a paper entitled `Towards an NLP-based Topic Characterization of Social Relations' presented at the 2012 ASE International Conference on Social Informatics (Washington D.C., USA, Dec. 14-16, 2012).}
\end{abstract}

\KEYWORD{keyword extraction; natural language processing; NLP; social network analysis; SNA; social computing; social relations; online communication.}

\maketitle

 \section{Introduction}

The surge in popularity of online social network services (SNS) such as Facebook, Twitter, and most recently Tumblr, is accompanied by a growing desire to identify popular, influential, or knowledgable users of these services. Social network analysis (SNA) provides a mathematical framework for these and related problems of social computing by defining a social network as a tuple $(V, E)$ with a finite set of vertices $V$, corresponding to social entities ({\em actors}), and a set of edges $E = \{\{x,y\} \mid x, y \in V \wedge x \neq y\} \subseteq \{\binom{V}{2}\}$, corresponding to social relationships among the entities. In other words, a social network is a graph induced by a binary relation in the mathematical sense, which is usually constructed by observing or modelling an actual social relation. In this work, we implement and evaluate a method of augmenting the social network graph with information extracted from communication between the actors, in order to move away from a purely topological approach to SNA.

\subsection{Problem Statement and Prior Research}
\label{sect:problemStatement}

Social network services allow their users to declare social relationships with other users. This can be bidirectional, requiring confirmation from both parties, or unidirectional, e.g. when ``following'' someone on Twitter. The set of declared relationships induces an explicit social network. Implicit social networks, also called {\em activity networks}, can be constructed by observing interactions between users and adding edges when a specific criterion is met, e.g. when the frequency of interaction within a period of time exceeds a threshold. In this way social networks can be extracted from any kind of online communication, as long as sender and recipient are identifiable. An example for this is the study of \cite{bird2006} on the extraction of social networks from collections of email messages, which lends empirical support to Wellman's \citeyearpar{wellman1997} earlier claim, that various kinds of ``electronic groups'' implicitly define social networks. Given that a typical social relationship is enacted by communication through various media, neither explicit nor implicit social networks from observations of a single medium will be an accurate representation of reality. The implicit network lacks relationships for which interaction mainly happens outside of the medium, while the explicit network overstates the importance of relationships that rarely engender any interaction at all. The latter observation is confirmed by a large-scale study of Facebook by \cite{viswanath2009}, who find that on average, less than 30\% of a user's relationships remain active from one month to the next. They explain the change in activity with temporal interaction patterns, and distinguish patterns of frequent and infrequent interaction. This observation mirrors Granovetter's \citeyearpar{granovetter1973} concept of tie strength and the different social roles of strong and weak ties. In conclusion, relationships among users of an SNS or comparable online community are conceptually heterogenous, and a substantial amount of information is lost when they are modelled by binary adjacency in a graph.

If it is possible to partition the set of relationships into homogenous subsets, one can construct one social network per subset, and analyse them jointly, as exemplified by \cite{louati2012}. However, this requires prior knowledge about the possible types of relationships, which is often not available. Instead, we introduce a weight function $f : E \rightarrow W$, where $W$ is the set of possible weights, and obtain a {\em weighted social network} $(V, E, f)$. \cite{barrat2004} demonstrate how to adapt standard methods of SNA to weighted networks with $W \subseteq \mathbb{R}$, and confirm that the combination of weights and graph topology affords new insights into the network structure. This model can be further generalised to weighted networks with $W \subseteq \mathbb{R}^n$. In a weighted social network, a relationship is treated as a point in a one- or multi-dimensional space. Usually, each dimension corresponds to an abstract concept from sociology or social psychology, e.g. the aforementioned tie strength. There are two ways of constructing a weighted social network:

\begin{itemize}
\item {\em Performing a survey among the actors.} This is done to obtain a reference or ``gold standard'' network for evaluation or for training of a supervised machine learning algorithm. In the most simple version, each actor is asked to directly rate his relationships on the previously specified scales, but more sophisticated interview techniques may be used.
\item {\em Extracting information from communication data.} Since the only observable part of a social relationship is the associated interaction via online services, communication data is the primary source of information. While some kinds of interaction could be classified as non-verbal, e.g. sharing a picture, we focus on verbal interaction, which is mostly limited to textual communication on current SNS. A common approach is to derive a low-dimensional representation of the relationship from statistics, such as frequency, average length, or degree of reciprocity, which are generated by aggregating communication meta-data. This places emphasis on the act of communication, while ignoring its content, the subjects of conversation. Extracting information from the textual content requires natural language processing (NLP).
\end{itemize}

In previous studies \citep{groh2011, hauffa2011}, we have identified two major problems with this approach: First, the assessment of a social relationship in terms of absolute numbers is hard for both human and computer. As a direct consequence, this complicates the generation of a reliable reference dataset. The quality of the dataset can be improved by use of sophisticated interview techniques or by increasing the data volume. Due to different positions in the social network, two actors will rarely have the same view on a relationship, so the expected lack of annotator agreement renders an improvement of data quality through multiple annotation unlikely.
Second, the performance of a predictor, especially in the unsupervised case, depends on the representation of the data, i.e. the choice of features. Using a set of heuristic features \citep[c.f.][]{groh2011} to predict emotional intensity and valence yielded unsatisfactory results, hinting at a complex latent structure.

To address these issues, we propose a representation of social relationships based on the content of communication. If all communication between two actors was available, one could build a comprehensive representation of what their relationship ``is about''. The content of the communication artefacts associated with a social relationship can be mathematically expressed as a vector space model \citep[VSM,][ch. 8.5.1]{manning1999}. Given a vocabulary of $n$ distinct words, a collection of documents is represented by an $n$-dimensional normalised vector, where each component $c_i$ is proportional to the frequency, within the collection, of the $i$-th word. To promote sparsity, we perform keyword extraction and keep only the top $k$ keywords with $k \ll n$. If $s_i$ is the score assigned to the $i$-th word by the keyword extraction algorithm, the $i$-th component of the vector is set to $s_i / k$ if the word is among the top $k$ keywords, and $0$ otherwise. Building a keyword-based representation does not involve any domain specific processing, and consequently it lacks a direct sociological interpretation. Yet the ubiquity of tag clouds in social media shows that sets of keywords can be effectively visualised. We claim that keywords are interpretable by humans, but also suitable for further processing by computer, and therefore are a useful intermediate representation of social relationships.

\subsection{Research Questions}
\label{sect:researchQuestions}

Based on the problem statement, we formulate two research questions:

\begin{enumerate}
\item It is possible to extract a set of keywords from communication artefacts that is considered a good representation of the associated social relationship by a human observer? The intuition is that given enough data, a pattern of keywords specific to the relationship will emerge from the background noise of words with mostly conversational function.
\item Will a representation considered accurate by a human observer conversely improve the performance of social computing systems?
\end{enumerate}

In this work, we focus on the first question and evaluate a keyword extraction system on text messages exchanged between users of the social network service Facebook. Section \ref{sect:keywordExtraction} describes the design of the keyword extraction system in detail. In section \ref{sect:dataAcquisition}, we build a Facebook application that analyses private messages and obtains user feedback on the quality of the extracted keywords. The results are presented in section \ref{sect:results}.

\section{Experimental Setting}

In order to obtain a representative sample of interpersonal communication, a large amount of messages has to be collected. We decided to use the social network service Facebook as a data source, because it is known for a large user base and network effects (``viral marketing''). Facebook allows users to mutually declare ``friendship'' and exchange private messages. The ``friend'' relation over the set of users corresponds to individual social relationships that may differ in emotional and geographical closeness. For example, Facebook is frequently used as the primary tool for maintenance of relationships between geographically distant persons \citep{bryant2009}.

Facebook offers an API that allows third-party developers to integrate their applications into the user interface and access user profile data as far as permitted by an individual user's privacy settings. The Facebook application developed for this study performs three distinct tasks: The private messages sent and received by a user are collected via the API, a set of keywords is extracted from the collected messages for each sender-recipient pair, and finally the user is asked to provide an assessment of the quality of the keywords.

\subsection{Keyword Extraction}
\label{sect:keywordExtraction}

The task of summarising one or more documents by selecting the most relevant words is known as keyword extraction. By formulating keyword extraction as a machine learning task, one can choose from supervised and unsupervised methods. Since the definition of relevance is domain specific, supervised learning in this setting requires a training corpus of messages, in which all keywords relevant to the relationship have been manually identified. However, reports on the construction of corpora annotated by similarly subjective criteria, e.g. the MPQA opinion corpus \citep{wiebe2005}, indicate potential problems with annotator agreement. To maximise agreement, a thorough definition of keyword relevance in a social context is necessary. Such a definition would have to be grounded in sociological theory, empirically validated, and formulated in a way that is comprehensible by the annotators. To the best of our knowledge, such a definition does not yet exist, so we limited the scope of this study to unsupervised methods. The basic steps of unsupervised keyword extraction are:
\begin{enumerate}
\item{\em Preprocessing} Tokenization and any per-token processing, e.g. part-of-speech tagging, that is required by the later steps.
\item{\em Candidate Selection} Statistically or linguistically motivated filtering to limit the set of candidates to words with a high probability of being relevant.
\item{\em Scoring and Ranking} Assignment of a numerical rating to the candidates, possibly removing low-ranked words from the candidate set.
\item{\em Keyphrase Formulation} Building coherent multi-word expressions from highly ranked words.
\end{enumerate}

Although a number of state-of-the-art keyword extraction systems are freely available, we decided to re-implement three algorithms from Hasan and Ng's \citeyearpar{hasan2010} survey to be able to adapt each part of the resulting system to the domain of online social interaction. A small number of preliminary experiments were conducted on data from individual relationships, with the purpose of tuning the system's implicit and explicit parameters. Each experiment was concluded by an interview with one of the involved actors. Our strategy was to start off with a language neutral keyword scoring algorithm and add pre- and post-processing steps as required.

In these experiments, straightforward implementations of the keyword extraction algorithms without any further processing performed poorly, not producing results that were interpretable in the context of social interaction. Linguistically motivated, and thus language-dependent, pre- and post-processing noticeably improved performance. Language identification is a requirement for any further linguistic processing. Before any actual processing, all messages not in English or German language are discarded by an n-gram based language classifier. The complexity of implementation rises with the number of languages to be supported, so we restrict the system to the languages most likely to be used by the participants, which are chosen in what amounts to an accidental sampling scheme. For each language a 3-gram classifier was trained on the Europarl corpus \citep{koehn2005}, with the added condition that the ratio of words to stop words is below an empirically chosen threshold of 4. Tokenization and part-of-speech tagging are performed by the Stanford PoS tagger \citep{toutanova2003} using the models {\em german-fast.tagger} and {\em left3words-wsj-0-18.tagger} for German and English respectively. PoS tags are used for candidate selection and as a means of word sense disambiguation when comparing words. \cite{gimpel2011} find that the Stanford PoS tagger, being trained on newswire text, performs worse when applied to Twitter messages. We experience a loss in performance consistent with the results of Gimpel et al., which can be attributed to stylistic features not present in the training corpus, mainly words containing punctuation characters (e.g. email addresses) and emoticons.

Candidate selection is performed by three filters: A PoS tag filter discards all words that are neither nouns nor adjectives, based on the results of \cite{mihalcea2004}. A stop word filter discards words that are known to be of little relevance. We augmented the stop word lists of the KEA project (http://www.nzdl.org/Kea) with domain specific vocabulary, such as abbreviations commonly found in text messages. Informal online communication often contains unique spelling variants of stop words that would receive an inappropriately high score by {\em idf} weighting and derived methods. To deal with these variants, stop words can be regular expressions, which are especially useful for filtering out elongated words and words where letters have been substituted with digits of similar shape. For example, ``(l|L)\textsuperscript{+}(o|O|0)\textsuperscript{+}(l|L)\textsuperscript{+}'' matches variants of ``lol'' (``laughing out loud''). Finally, a set of heuristics discards very long words ($\ge 30$ characters), words with a ratio of length to number of unique letters that exceeds a threshold of 3, and words containing punctuation characters commonly used as part of emoticons. The remaining candidate words are reduced to their word stems by applying the algorithms provided by the Snowball library (http://snowball.tartarus.org). For English, this is Porter's stemming algorithm. From this point on, two words are considered equal if their word stems and PoS tags are the same. The original word forms are kept for displaying the word to the user.

Three methods of keyword scoring have been implemented, {\em tf-idf} as described by \cite{manning1999}, TextRank \citep{mihalcea2004}, and a custom variant of TextRank that operates on directed graphs. TextRank is essentially an application of PageRank to a graph constructed by treating words as vertices and adding an edge between two words if they co-occur within a specified distance (``window size''). As text messages often lack proper punctuation, and a conversation may be spread across multiple messages, we deviate from the original algorithm by adding edges for words even if they are separated by a sentence or message boundary. TextRank is applied to an undirected and unweighted graph constructed with a window size of 2. The damping factor is set to $0.85$ and the convergence threshold is $10^{-5}$. We also evaluate a variant of TextRank that operates on a directed graph generated according to the rules set forth by \cite{litvak2011}. Directed edges are added between words that occur in direct succession. For both graph based methods the maximum number of iterations is set to 100, to put an upper bound on processing time.

The main difference between {\em tf-idf} and TextRank is that the former judges a word from its frequency in a reference corpus in addition to its frequency in the document, so that the composition of the reference corpus has to be considered a parameter. In the present implementation the reference corpus consists of all messages sent or received by the current user, excluding those associated with the currently examined relationship. The {\em idf} weight of a word $w$ is computed as $\log(D / (1 + df_w))$, where $D$ is the number of threads in the corpus, while the document frequency $df_w$ is the number of individual messages containing the word. This adaption of the original {\em idf} weighting scheme is due to the observation that the chain of replies, which Facebook collects as a thread, may contain a large number of short messages on the same subject. For performance reasons, PoS tags are not taken into account when computing the document frequency. Since the graph based methods do not originally incorporate an {\em idf} weighting scheme, preliminary experiments were performed with an additional post-processing step of adjusting the score of a word according to its document frequency: If the ratio of document frequency to word length exceeds a threshold of 3, the score is discounted. This filter targets short words that occur frequently in the whole corpus. Unexpectedly, this step consistently improved perceived keyword quality, even in conjunction with {\em tf-idf} scoring.

Some concepts cannot be appropriately represented by single words, e.g. place names such as ``New York''. Therefore in a final processing step, words that occur next to each other in the original document are combined to form a single keyphrase under certain circumstances: First, a list of all sequences of candidate words that occur in the messages is complied. The score of a sequence is the harmonic mean of the scores of those constituent words that in the previous step passed through the {\em idf} post-processing filter. Sequences containing less than two such words and sequences that occur as parts of longer sequences are only considered if the requested number of keywords cannot be generated by other means. Using the harmonic mean for computing the score ensures that keyphrases are only constructed from words that are good keywords on their own, which effectively penalises longer phrases. From here on, we use the term {\em keyword} to refer to both single words and multi-word expressions / phrases.

The result of scoring and keyphrase formation is list of keywords and phrases ordered by score. From this list, the $n$ highest ranked entities are chosen as a representation of the set of exchanged messages and thus of the relationship as a whole. This final selection can also be performed by setting a score threshold, which is effectively a lower bound on relevance. Multi-word expressions are displayed as they occur in the original message. Word stems that correspond to multiple different word forms in the original messages are represented by the word form with the lowest Levenshtein distance to the stem.

Keyword extraction suffers from a problem known as the {\em vocabulary gap} \citep{liu2012}, where none or very few of the appropriate keywords occur in the document itself. This problem appears to be highly domain specific: \cite{turney1999} performs keyword extraction on an email corpus and finds that on average 97.9\% of keywords proposed by human annotators occur in the text, compared to 65.3\% for a corpus of web pages. Since online communication was found to be sufficiently self-contained, we do not specifically address the vocabulary gap.

Keyword extraction systems are usually evaluated by comparing the set of extracted keywords to a manually compiled reference set in terms of precision and recall. However, \cite{turney2003} remarks that a particular document might be represented equally well by more than one set of keywords, and recommends having the output of the system rated by human judges. This is consistent with the results of \cite{jones2001}, who find a statistically significant agreement on the quality of keywords between different human assessors. \cite{barker2000} attribute this effect to keyword coherence: ``Judges did not prefer keyphrase sets based simply on the individual keyphrases they contained. A set of keyphrases is somehow more than the sum of its individual keyphrases.'' Furthermore, previous studies \citep{hauffa2011} indicate that generating an appropriate set of reference keywords will be difficult: People asked to come up with ``tags'' for a relationship will focus on functional aspects rather than conversation topics, favouring tags corresponding to  emotional intensity (``good friend'' vs. ``best friend'') or the social setting (``high school''). For these reasons we decided to evaluate the system by presenting the participants with the results of different methods of keyword extraction and asking them to rate the quality of each set of keywords on a numeric scale. Note that our evaluation setting differs from Barker and Cornacchia's setting in three key points:
\begin{itemize}
\item The corpus of each participant consists of a unique subset of his own communication artefacts instead of a single previously chosen set of documents.
\item The participants are specifically asked to judge the keywords as a representation of a social relationship instead of judging how well they summarise the content of the messages.
\item The visual presentation (order and font size) of the keywords, corresponding to the relevance score, is to be considered in the judgement.
\end{itemize}

\subsection{Data Acquisition}
\label{sect:dataAcquisition}

\begin{figure*}
\caption{The keyword assessment screen}
\label{figure:ratingScreen}
\centerline{\includegraphics[width=2.8in]{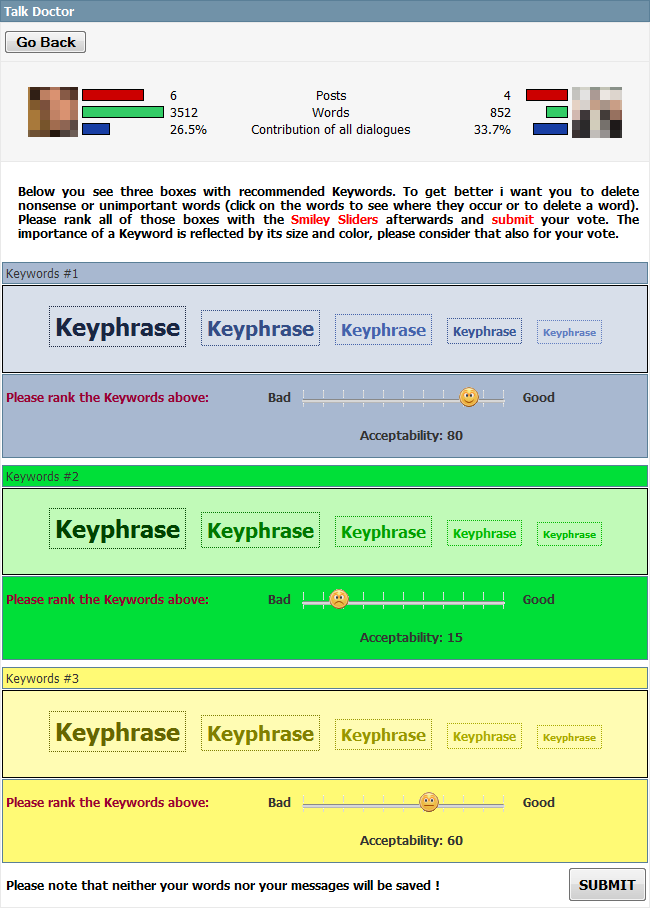}}
\end{figure*}

The ``Talk Doctor'' application is designed as a virtual advisor that provides the user with communication statistics as an incentive to use the application and recommend it to other users. Like the underlying keyword extraction system, the user interface is bilingual (German and English). The language is chosen according to the user profile settings. After launching the application, the first screen describes the setting of the survey: ``Imagine you're trying to represent the relationship between two people by a few keywords.'' The following screen lists the user's contacts by name, sorted by message count to emphasise the most salient contacts. When the user selects a contact, keyword extraction is performed on the messages exchanged between the user and the chosen contact. Finally, the screen shown in figure \ref{figure:ratingScreen} appears, visualising the three sets of keywords generated by applying the scoring algorithms described in section \ref{sect:keywordExtraction}. We chose to generate 20 keywords for each contact to keep the visualisation simple and not to overly strain the attention span of the participant. For a vector space model as described in section \ref{sect:problemStatement}, a higher amount of keywords might be preferable. Keywords are displayed in order of their relevance score, which is also reflected by the font size. By clicking a keyword, the user can view it in the context of the messages in which it occurs. Keywords deemed completely irrelevant can be deleted, but at least one keyword has to be left. Located below each set of keywords is a slider for rating the quality of the set on a scale of 0 to 100. The slider's handle is a smiley face that changes its expression while being moved. A rating of 100 means that each of the keywords contributes to an accurate representation of the content of the relationship. Lower values indicate a higher proportion of unrelated keywords or a generally lower quality of representation. In terms of information retrieval, this is closer to precision than to recall, reflecting the expectation that most relationships can only be partially observed through online communication.

Messages are collected from inbox and outbox without distinction, so the extracted keywords pertain to an undirected relationship edge. If a sufficient number of messages is available for each direction, keywords for directed edges can be generated by processing incoming and outgoing messages separately. Messages with more than one recipient are discarded to ensure a constant level of significance for the relationship. If the number of suitable messages is insufficient, the user is notified and asked to choose another contact. Even though the API offers access to communication via Facebook's instant messaging system, such messages are ignored due to their conceptually different nature.

Measures are taken to protect the participants' privacy: Facebook requires that each application using the API for accessing private messages is submitted for review and whitelisting. When installing the application, the user has to explicitly grant access to private messages. All communication between the application server and Facebook takes place over TLS encrypted channels. The ``Talk Doctor'' application does not retain copies of the messages after processing, and the extracted keywords are encrypted by a MD5 one-way hash function before being stored.  The hashes are used in place of the original words in all further processing and evaluation. While the hashing is crucial for preserving privacy, it also limits evaluation and visualisation of the results, as discussed in section \ref{sect:results}.

The initial set of participants was recruited from the acquaintances of the authors and by advertising on university noticeboards and topically appropriate Internet discussion boards. To achieve the goal of distribution by word of mouth, participants were requested to publicise the application among their Facebook contacts. As an additional incentive, all participants were entered in a raffle for a portable media player.

\section{Results}
\label{sect:results}

Data was collected over a period of approximately 2.5 months. During that period, 98 Facebook users installed the application, and 71 users actually submitted usable data. Assessments were submitted for 275 relationships. 46 users (65\%) on average removed one or more words from a keyword set before submitting their assessment. The average number of messages associated with a relationship is 20, with an average number of 36 words per message. There is a strong linear correlation between the amount of messages sent and received per relationship, with Pearson's correlation coefficient $\rho = 0.92$ for the ratio of messages and $\rho = 0.9$ for the ratio of the overall word counts. This is evidence for reciprocal messaging behaviour on Facebook.

\begin{table*}
\caption{Human quality assessment of the keyword extraction system (in percent of a perfect score)}
\label{table:results}
\begin{tabular}{llll}
& \begin{tabular}[t]{@{}c@{}}TextRank\\(undirected)\end{tabular} & \begin{tabular}[t]{@{}c@{}}TextRank\\(directed)\end{tabular} & {\em tf-idf} \\
\toprule
avg. per relationship & 72.43 \% & 67.76 \% & 72.07 \% \\
avg. per user & 69.86 \% & 65.20 \% & 69.73 \% \\
std.dev. per relationship & 19.88 & 20.13 & 21.56 \\
std.dev. per user & 17.51 & 18.04 & 20.18 \\
\end{tabular}
\end{table*}

\begin{figure*}
\caption{Histograms of the human assessments of keyword sets for individual relationships}
\label{fig:histograms}
\centerline{
\subfloat[TextRank (undirected)]{\includegraphics[width=1.75in]{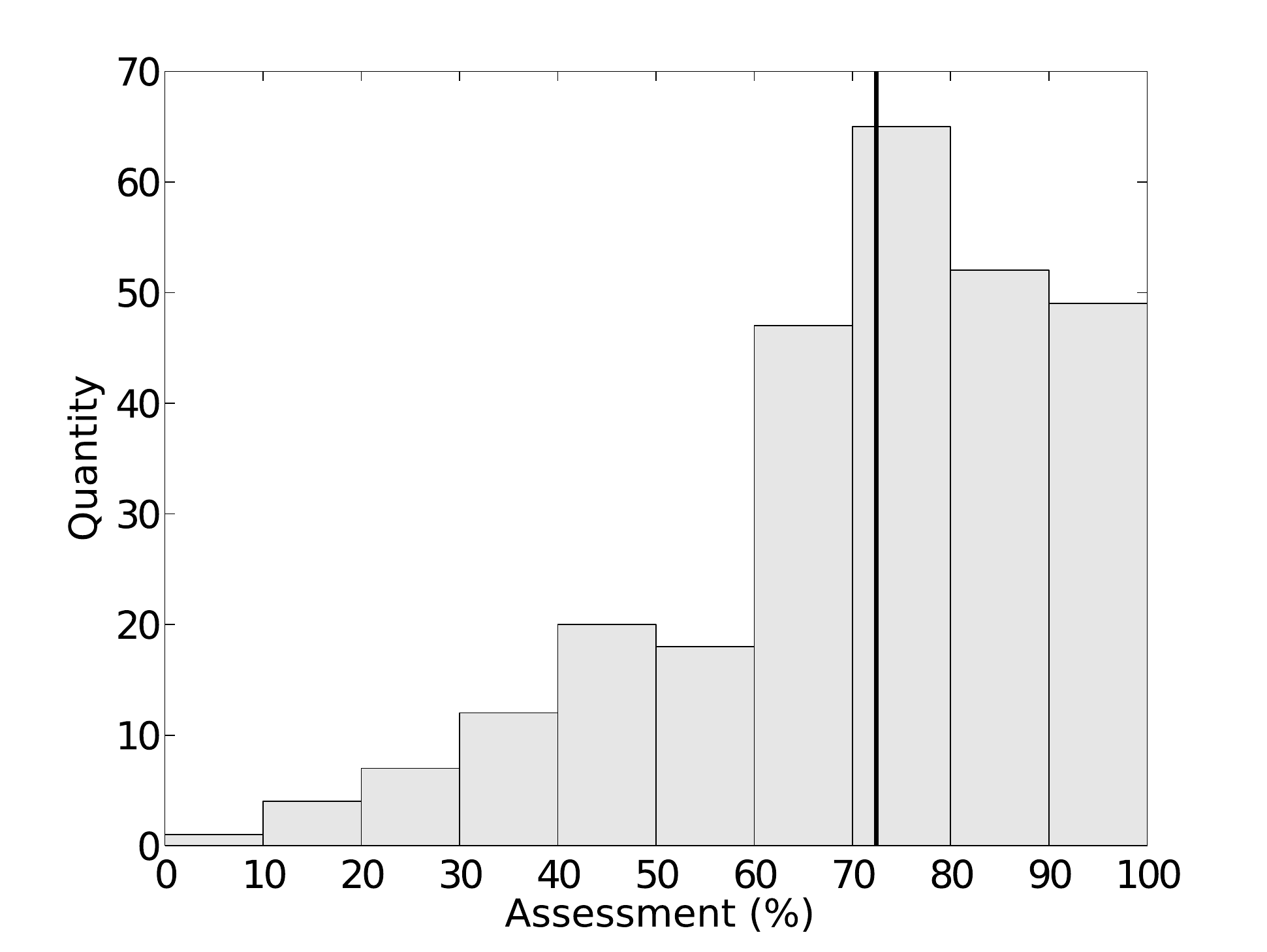}
\label{fig:histograms:tru}}
\hfil
\subfloat[TextRank (directed)]{\includegraphics[width=1.75in]{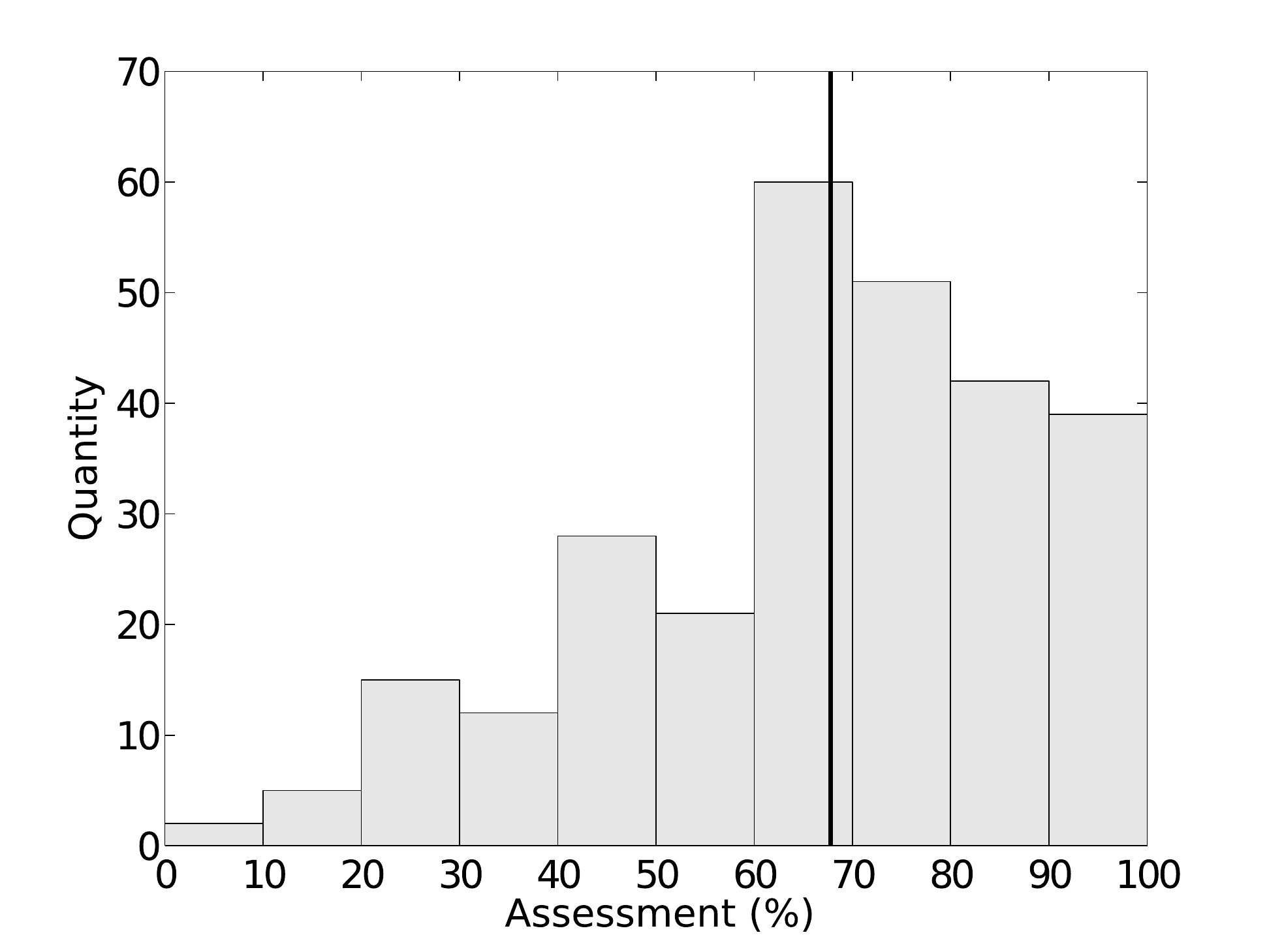}
\label{fig:histograms:trd}}
\hfil
\subfloat[\em tf-idf]{\includegraphics[width=1.75in]{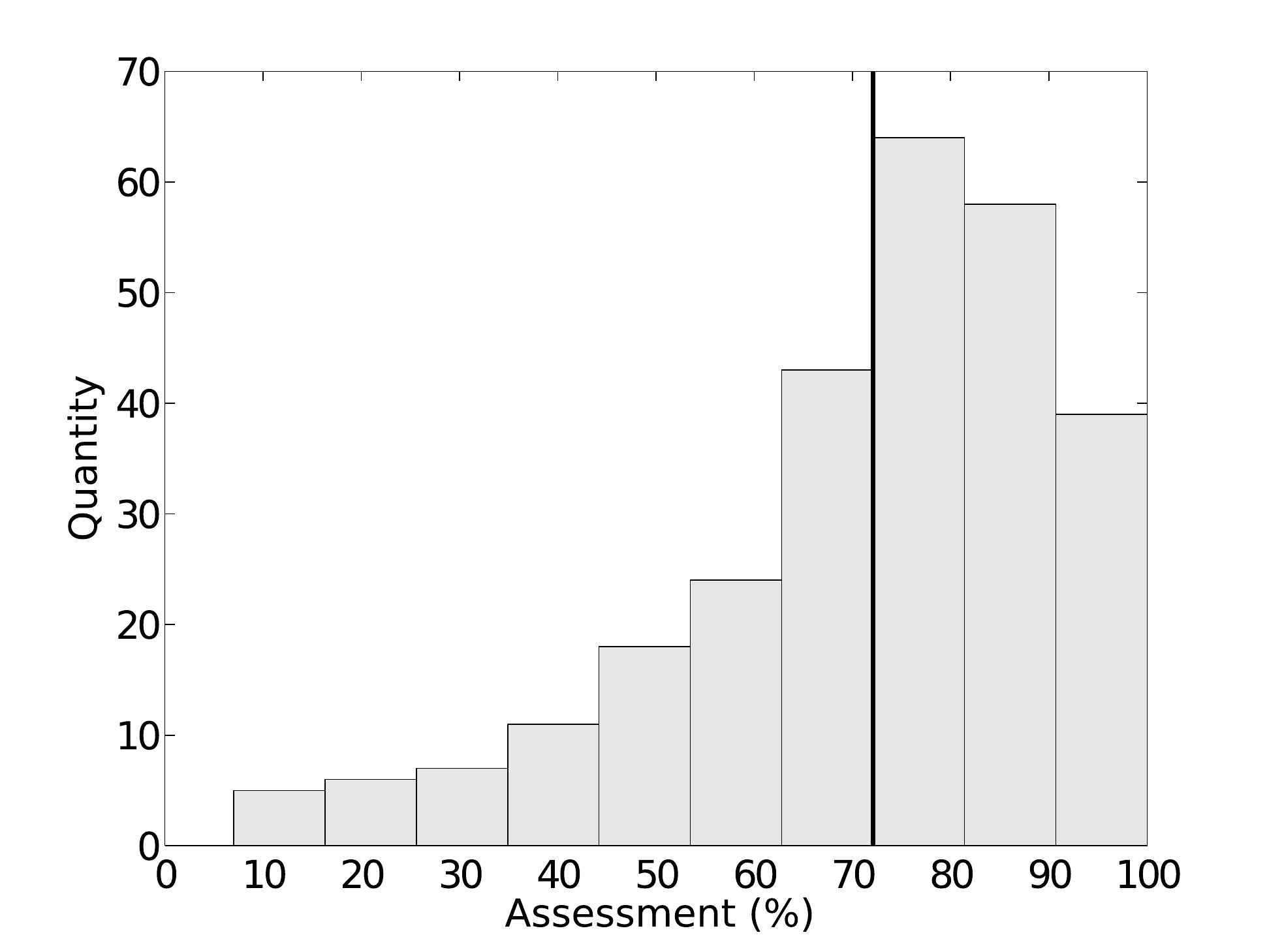}
\label{fig:histograms:tfidf}}
}
\end{figure*}

\begin{figure*}
\caption{Histograms of the human assessments of keyword sets aggregated by user}
\label{fig:histograms_u}
\centerline{
\subfloat[TextRank (undirected)]{\includegraphics[width=1.75in]{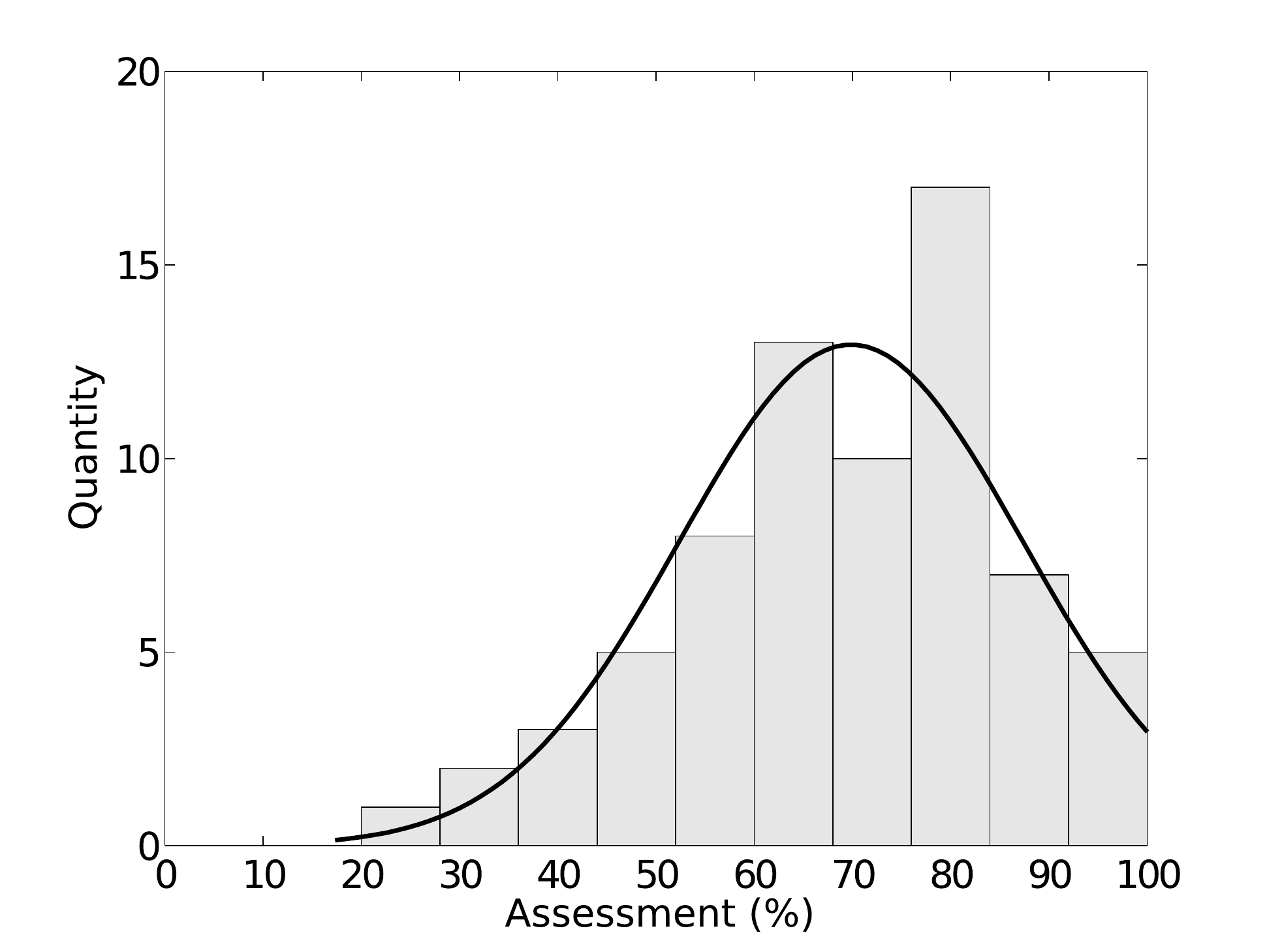}
\label{fig:histograms:tru_u}}
\hfil
\subfloat[TextRank (directed)]{\includegraphics[width=1.75in]{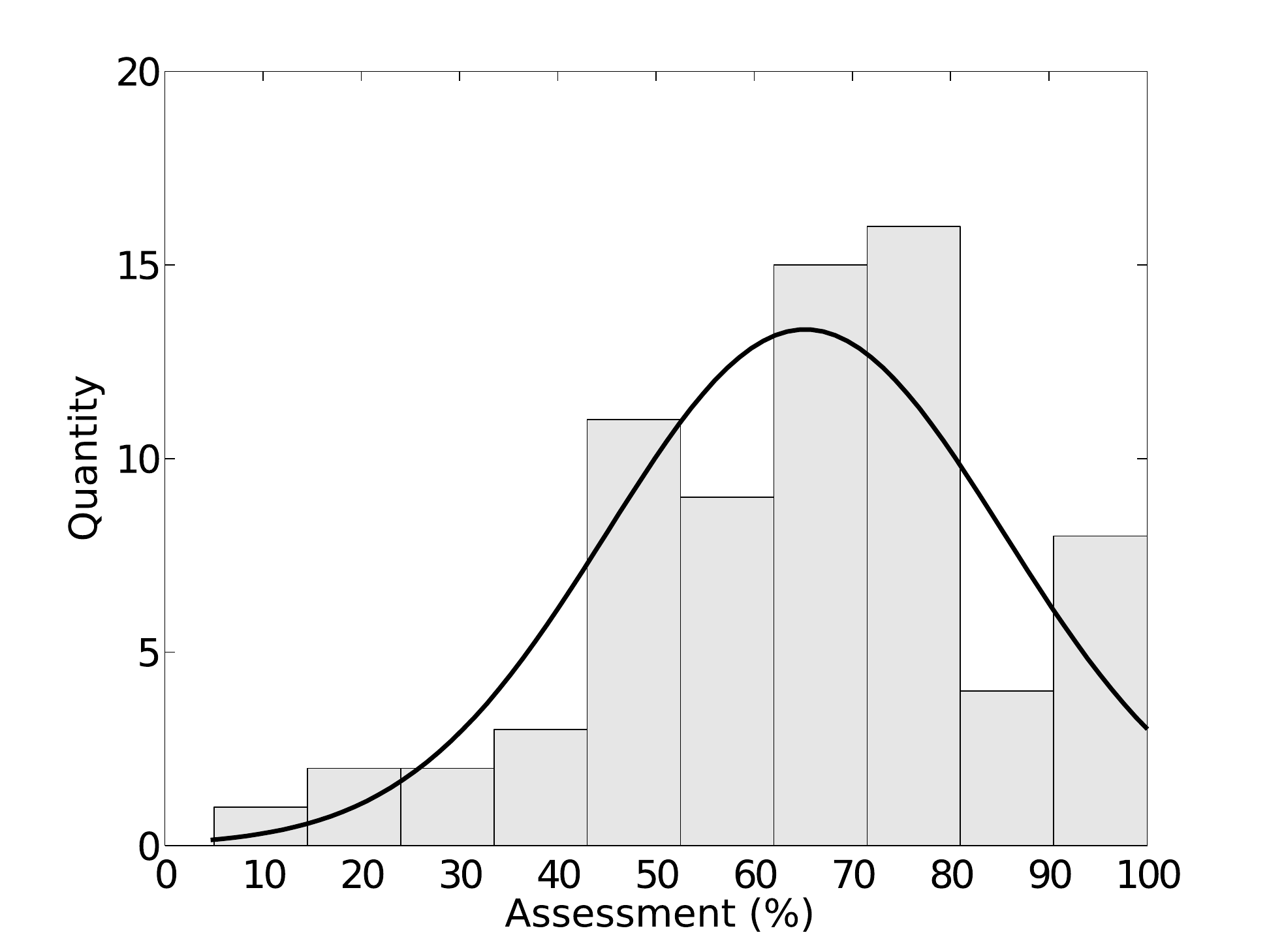}
\label{fig:histograms:trd_u}}
\hfil
\subfloat[\em tf-idf]{\includegraphics[width=1.75in]{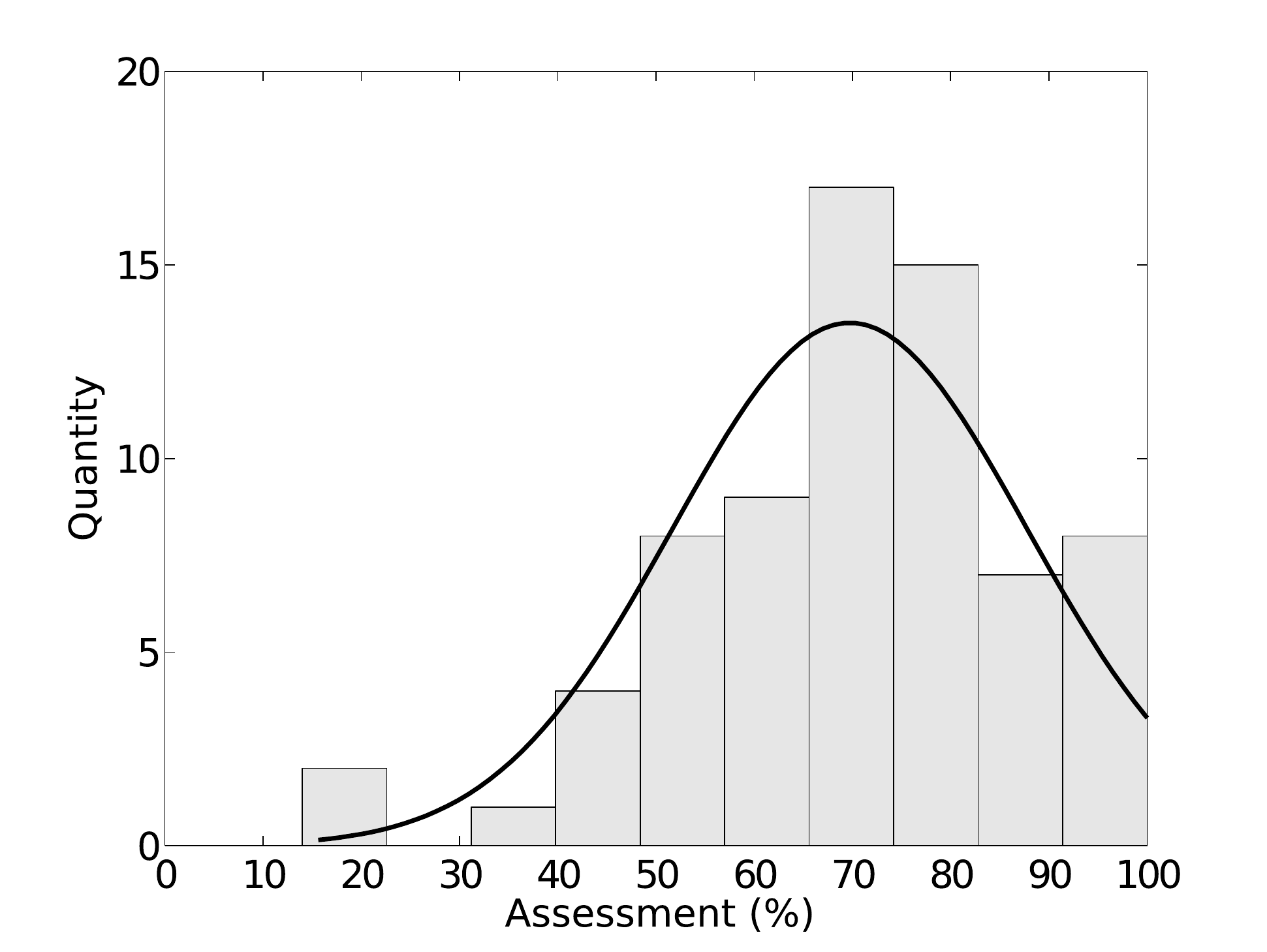}
\label{fig:histograms:tfidf_u}}
}
\end{figure*}

The main results of the study are summarised in table \ref{table:results}. The upper half of the table contains empirical estimates of mean and standard deviation of the assessments. Each user may submit assessments for multiple relationships, so the resulting data can be interpreted in two ways: Considering each relationship as an individual entity may introduce bias caused by users who submitted a high number of assessments, but makes best use of the available data. Taking the average of all assessments submitted by a user avoids bias but reduces the dataset to 26\% of its original size. The assumption of normality is tested for both the unmodified ``per relationship'' data set and the ``per user'' average using the Lilliefors and $\chi^2$ tests with $\alpha = 0.05$. A normal distribution fitted to the data is compared to the distribution generated by a Gaussian kernel density estimator. The tests are in agreement for all three methods of keyword scoring: The tests fail to reject the null hypothesis for the ``per user'' data, thus providing evidence for a normal distribution. The tests reject the null hypothesis with $p < 0.01$ for the ``per relationship'' data, indicating a significant deviation from the normal distribution. When interpreting these counterintuitive results, one has to take the overall low sample size into account. Figures \ref{fig:histograms} and \ref{fig:histograms_u} show the histogram of assessments for each method of keyword scoring. The histograms in figure \ref{fig:histograms} are computed from the individual assessments, and the superimposed vertical lines represent the mean assessment. The histograms in figure \ref{fig:histograms_u} are computed from the average assessment scores of each user and are superimposed with the density of the fitted normal distribution.

No meaningful correlation could be found between the assessment and the conversation activity in terms of word or message count. For all three methods of keyword scoring, the magnitude of correlation $\rho^2$ is below $0.04$, except for {\em tf-idf}, where the correlation between per-user average assessment and message count is $\rho = -0.33$. This may be caused by the domain specific {\em idf} formulation, which incorporates the message count.

\subsection{Deleted Keywords}

Participants are encouraged to delete those keywords they consider completely irrelevant. An analysis of the deletion behaviour of the participants may help to identify opportunities to improve the keyword extraction process. This is where we suffer most from the keyword hashing, which completely obscures the semantics of the keywords.

\begin{table*}
\caption{Statistics of keyword deletion behaviour}
\label{table:deletion}
\begin{tabular}{llll}
& \begin{tabular}[t]{@{}c@{}}TextRank\\(undirected)\end{tabular} & \begin{tabular}[t]{@{}c@{}}TextRank\\(directed)\end{tabular} & {\em tf-idf} \\
\toprule
users deleting $\ge 1$ word & 46 & 46 & 45 \\
avg. words deleted & 3.6 & 2.9 & 3.4 \\
$\rho$ per relationship & -0.79 & -0.38 & -0.85 \\
$\rho$ per user & 0.19 & 0.32 & -0.64 \\
\end{tabular}
\end{table*}

According to the usage statistics in table \ref{table:deletion}, more than one third of the participants did not make use of the facility for removing irrelevant keywords, while those who did, removed 3.3 keywords on average. Investigating the reasons for this dichotomous behaviour, we construct two different hypotheses by looking at the correlation between the quality assessment and the number of removed keywords per relationship and per user. The first hypothesis proposes that there is a subset of relationships for which the keyword extraction system is able to generate high quality keywords, so that no removal is necessary. A strong negative correlation on the level of individual relationships would be consistent with this hypothesis. The second hypothesis is that a subset of users consistently ignores the possibility to discard low quality keywords. In this case, one would expect the correlation between per-user average and number of removals to be weak. As shown in table \ref{table:deletion}, the results are different for each method of keyword scoring: In the case of undirected TextRank, a strong negative correlation per relationship and a low correlation per user can be seen as evidence for either hypothesis. For directed TextRank, both correlations are low, which points towards the second hypothesis and indicates a generally lower quality of the generated keyword sets. For {\em tf-idf}, both correlations are negative, indicating that more users remove keywords. The combined evidence points towards a subset of users that does not delete keywords. In this case, a strong negative correlation, as can be found for {\em tf-idf}, might also indicate that the perceived quality of a set of keywords frequently did not improve even after removing the most irrelevant keywords.

Comparing the set of words discounted by the {\em idf} filter to the set of words deleted by more than one user, we find an overlap of 45\% for the German language. This agrees with the preliminary experiments of section \ref{sect:keywordExtraction}, where the filter was found to improve the perceived quality. Results for English are not representative due to lack of data.

\begin{table*}
\caption{Part-of-Speech patterns of deleted words / phrases}
\label{table:pos}
\begin{tabular}{lrr}
pattern & \% retained & \% removed \\
\toprule
N & 47.80 \% & 33.90 \% \\
A\textsuperscript{+}N\textsuperscript{+} & 11.70 \% & 10.80 \% \\
N\textsuperscript{+}A\textsuperscript{+}N\textsuperscript{+} & 0.30 \% & 0.18 \% \\
N\textsuperscript{+}A\textsuperscript{+}N\textsuperscript{+}A\textsuperscript{+}N\textsuperscript{+} & 0.02 \% & 0.00 \% \\
\midrule
A\textsuperscript{+}N\textsuperscript{+}A\textsuperscript{+}N\textsuperscript{+} & 0.10 \% & 0.18 \% \\
other & 0.08 \% & 0.19 \% \\
A\textsuperscript{+}N\textsuperscript{+}A\textsuperscript{+} & 0.20 \% & 0.35 \% \\
N\textsuperscript{+} & 4.40 \% & 6.00 \% \\
N\textsuperscript{+}A\textsuperscript{+} & 1.60 \% & 3.70 \% \\
A\textsuperscript{+} & 4.40 \% & 6.90 \% \\
A & 29.40 \% & 37.80 \% \\
\end{tabular}
\end{table*}

Table \ref{table:pos} breaks down the global sets of retained and removed keywords by part-of-speech tag. The table is sorted by ratio of retained to removed keywords. The letter `N' denotes nouns, `A' adjectives, and a plus sign after a letter matches one or more words of the specified type. Single nouns and adjective-noun combinations ending in a noun are more likely to be kept, while ungrammatical fragments ending in an adjective are more likely to be rejected. Contrary to the findings of \cite{mihalcea2004}, single adjectives are also frequently deleted. From an NLP perspective, noun phrases appear to be the most salient sources of keywords.

\section{Related Work}

\cite{culotta2004} build a system for extracting social networks from the web that acts like a web crawler. The system extracts keywords from the discovered personal websites for the stated purpose of expertise identification. The keyword quality is not evaluated beyond basic visual inspection. \cite{mori2007} investigate the extraction of relations between persons and entities from the web using a search engine and specially crafted queries. The individual relationships are annotated with highly ranked keywords from the search results. The keywords are used for clustering the social relationships, and the clusters are evaluated against a manually labeled reference dataset. \cite{liu2012} combine conventional keyword extraction with techniques from machine translation to address the vocabulary gap problem. They generate keywords for the nodes of the social network of a microblogging service, and provide a visualisation to the users via a web application. Though they do not evaluate the keyword extraction, they report success in drawing a large number of users to their application, which indicates that online services for public and semi-public communication are a better target for this kind of study than closed systems like Facebook. What sets our study apart from the aforementioned works is the evaluation of whole keyword sets by human judges according to a specific measure of quality.

Topic models provide a principled alternative to heuristics for scoring keywords, such as {\em tf-idf}. The ART model of \cite{mccallum2007} is a specific topic model for electronic communication. A topic model expresses documents as multinomial probability distributions over topics, which are in turn distributions over words. The parameter vector of the per-document distribution can be directly assigned to a social network edge as a weight, but generating a set of keywords from a topic model is possible as well \citep{zhao2011}. \cite{chang2009a} discuss the evaluation of topic models and the associated issues, which are similar to those of keyword extraction systems.

\section{Discussion}

A Facebook application has been developed to evaluate a keyword extraction system that operates on communication artefacts. The goal was to demonstrate its ability to generate keyword sets that are perceived as carrying information about social relationships by human judges. The keywords are supposed to augment the structural information of a social network graph.

The attempt to mobilise a large user base by word of mouth marketing did not succeed, seeing that the user base did not significantly grow beyond the initial set of participants. Observations during the recruitment process indicate that privacy concerns among the prospective participants are one reason for the slow adoption.

As demonstrated in section \ref{sect:keywordExtraction}, the basic keyword extraction algorithms require extensive domain specific tuning to produce high quality results when applied to online communication. This may be attributed to characteristics of the text type: Short text messages exchanged in a casual environment are often noisy in terms of spelling and grammar. Achieving domain adaption by formulating heuristics and tuning their parameters is a resource intensive process, and the results are not necessarily transferable to other computer aided communication settings, even ones that are highly similar in concept. The problem is exacerbated by the introduction of linguistic processing steps, which require training and customisation for each language that is to be supported. None of the keyword scoring methods discussed here are part of a principled framework that would allow for controlled domain adaption and tuning. Such a framework would have to be able to account for the individual characteristics of communication data from different sources, e.g. limited visibility of the social network graph, specific use of language, uni-directionality (email inboxes), one-to-many communication (Twitter), or limited access to historical data. The performance improvement obtained by filtering the candidate words according to their inverse document frequency suggests an extensional approach that derives $idf$-like information from a reference corpus.

The results of the keyword quality assessment are encouraging and show potential for further development. A subjective quality assessment of about 70\% indicates that the participants did indeed see value in the selection and presentation of keywords. Given that a social relationship is only partly observable through communication artefacts, even less so if the analysis is restricted to direct communication between the two actors, a comprehensive textual representation is unattainable in most cases. The deletion behaviour confirms the potential for improvement: Low quality ``noise'' keywords are one reason for bad ratings, but do not fully explain the observations. Some relationships appear to be harder to express through keywords, either intrinsically or due to a lack of data.

Looking at the individual scoring algorithms, directed TextRank yields the worst results. It exploits word order information and is thus more susceptible to ``noisy'' text. {\em Tf-idf} favours keywords that set a message apart from others, but keywords that connect different messages in a thread may be important as well, which possibly explains the slightly better performance of undirected TextRank. However, {\em tf-idf} might benefit from a larger corpus, which according to \cite{paukkeri2008} should be constructed to represent all aspects of language use. An unexpected difference between TextRank and {\em tf-idf} is in the way the removal of keywords by the participants affects the perceived quality of the keyword set: In the case of TextRank, there is evidence for an overall positive effect of keyword removal, while in the case of {\em tf-idf} the effect appears to be much less pronounced. Due to low sample size it is not possible to definitely recommend one algorithm over the others.

Future research will focus on improving the perceived quality of keywords and on making an attempt to answer the research question posed in section \ref{sect:researchQuestions}, which means demonstrating the value of keywords as an intermediate representation of social relationships for computational tasks. Investigation into alternative methods of obtaining a low-dimensional representation of communication data, including topic models, for use within the weighted social network model, might also prove fruitful.

\bibliography{paper}
\bibliographystyle{hapalike}

\end{document}